\documentclass[12pt,preprint]{aastex}

\newcommand{\COJ}[2]{\mbox{CO$\,J={#1} - {#2}$}}
\newcommand{\J}[2]{\mbox{$J={#1} - {#2}$}}
\newcommand{\JJ}[2]{\mbox{${#1} \rightarrow {#2}$}}
\newcommand{\LOJ}[2]{\mbox{L$'_{\rm{CO}}({#1} - {#2})$}}
\newcommand{\arcsecs}{\mbox{$^{\prime\prime}$}}
\newcommand{\Msolar}{\mbox{$M_{\odot}\,$}}
\newcommand{\Lsolar}{\mbox{$L_{\odot}\,$}}
\newcommand{\degs}{\mbox{$^{o}$}}              
\newcommand{\XCO}{\mbox{$X_{\rm{CO}}$}}              

\shorttitle{Gas and Dust in the Extremely Red Object ERO\,J164502+4626.4}
\shortauthors{}

\begin{document}


\title{Gas and Dust in the Extremely Red Object ERO\,J164502+4626.4}


\author{Thomas R. Greve}
\affil{Institute for Astronomy, University of Edinburgh, Blackford Hill, Edinburgh EH9 3HJ, United Kingdom}
\email{tgreve@roe.ac.uk}
\author{Rob J. Ivison}
\affil{UK ATC, Royal Observatory, Blackford Hill, Edinburgh EH9 3HJ, United Kingdom}
\email{rji@roe.ac.uk}
\author{Padeli P. Papadopoulos}
\affil{Department of Physics \& Astronomy, University College London, Gower Street, London WC1E 6BT, United Kingdom, and Sterrewacht Leiden,  P. O. Box 9513,  2300 RA Leiden, The Netherlands}
\email{pp@star.ucl.ac.uk}



\begin{abstract}
We report the first detection of the lowest CO transition in a sub-millimetre 
bright galaxy and extremely red object (ERO) at $z=1.44$ using the Very Large 
Array\footnote{The Very Large Array (VLA) is operated 
by the National Radio Observatory, which is a facility of the National Science
Foundation, operated under cooperative agreement by Associated Universities, 
Inc.}.  
The total \J{1}{0} line luminosity of ERO\,J164502+4626.4 
is $(7\pm 1) \times 10^{10}$\,K\,km\,s$^{-1}$\, pc$^2$, which yields a total
molecular gas mass of $\sim 6\times 10^{10}\,\Msolar$.
We also present a map of the 850-$\mu$m continuum emission 
obtained using SCUBA, from which we infer 
a far-IR luminosity and dust mass of $L_{FIR} \sim 9 \times 10^{12}\,\Lsolar$ 
and $M_d \sim 9 \times 10^{8}\,\Msolar$.
We find tentative evidence that the CO and sub-mm dust emission is extended 
over several tens of kpc. If confirmed by high-resolution imaging, this 
implies that ERO\,J164502+4626.4 is not simply a high redshift counterpart of 
a typical Ultra Luminous Infrared Galaxy (ULIRG).
\end{abstract}


\keywords{galaxies: structure -- galaxies: individual: (ERO\,J164502+4626.4) -- galaxies: ISM}


\section{Introduction}
Observations of CO provide one of the most powerful methods of probing the
interstellar medium (ISM) in galaxies -- i.e.\ determining the amount of
molecular gas available to fuel star formation and accretion onto active
galactic nuclei (AGN), the two processes believed to generate the large
far-infrared (far-IR) luminosities of star-forming galaxies
\citep{Sanders-et-al-1991}. The first detection of CO at high redshift
 \citep{Brown-and-vanden-Bout-1991} revealed its potential to trace
metal-enriched molecular gas in the early Universe. CO is thus key to our
understanding of galaxy formation and evolution. 

Extremely red objects (EROs - usually defined as galaxies with $R-K \ga 5.3$, e.g.~Moriondo, Cimatti \& Daddi 2000)
constitute a bi-modal population - a combination of dusty,
starbursting systems \citep{Dey-et-al-1999} and evolved ellipticals \citep{Dunlop-et-al-1996}. 
While both types
of objects appear similar in the optical/near-IR, the detection of sub-millimetre
continuum emission and/or CO unambiguously pinpoints the dusty EROs.
A significant fraction of the sub-mm-selected
population of high-redshift dust-enshrouded starburst galaxies
is believed to be associated with EROs \citep{Smail-et-al-1999}.
Since the advent of large-format bolometer arrays such as
SCUBA \citep{Holland-et-al-1999} and MAMBO \citep{Kreysa-et-al-1998}, over a hundred 
sources have been detected in (sub)millimetre surveys 
\citep{Smail-et-al-1997, Hughes-et-al-1998,Bertoldi-et-al-2001}.
In cases where optical/IR counterparts of sub-mm galaxies are available, it is found that
the sources typically are distorted, multi-component systems with one or more components
being an ERO \citep{Ivison-et-al-2002}.
The sub-mm population has a median redshift of $z \simeq 2.4$ \citep{Chapman-et-al-2003},
and is widely believed to comprise the progenitors of present-day spheroids and massive ellipticals.
At present, however, the molecular gas content of sub-mm selected galaxies remains largely
unknown, and a systematic inventory of the molecular gas and its properties in these
sources is required in order to properly address the question of their typical mass and
evolutionary status.
Of particular importance is the
ratio of gas and dynamical masses in such objects, since it can help determine
whether sub-mm galaxies are merely high-redshift analogues of local ultra luminous 
infra-red galaxies (ULIRGs) or massive, large-scale galaxy-formation events.
The difficulties arise mainly because sub-mm selected galaxies have
proved extremely difficult to identify in the optical 
(cf.~Ivison et al.~1998, 2000). As a result only
a handful of sub-mm selected galaxies have been detected in CO to date
\citep{Frayer-et-al-1998,Frayer-et-al-1999,Ivison-et-al-2001,
Downes-and-Solomon-2003,Genzel-et-al-2003,Neri-et-al-2003}

ERO\,J164502+4626.4 \citep{Graham-and-Dey-1996} was amongst the first EROs discovered by \citet{Hu-and-Ridgway-1994}.
Deep near-IR and optical spectroscopy of
J164502 put it at a redshift of $z=1.44$ \citep{Graham-and-Dey-1996,Dey-et-al-1999},
and the detection of the H$\alpha$ line and the
[O\,II]$\lambda$3726,3729 doublet in its spectrum suggested that J164502 is an
actively star-forming galaxy, possibly containing an AGN, and not an evolved
elliptical at $z=2-4$ as initially suggested by \citet{Hu-and-Ridgway-1994}. 
The presence of large amounts of dust was inferred from sub-mm observations by \citet{Cimatti-et-al-1998}
and \citet{Dey-et-al-1999}, the latter detecting J164502 in the continuum at 450, 850 and
1350\,$\mu$m. These observations together with detections of \COJ{2}{1} and \J{5}{4}
by \citet{Andreani-et-al-2000}, which revealed the presence of large quantities of molecular gas,
unamibiguously demonstrated that J164502 is a gas-rich dust-enshrouded
starburst galaxy.

Here we present observations of the \COJ{1}{0} emission from 
J164502 using the VLA, providing the first detection of the lowest,
and thus least excitation biased, CO line in a galaxy typical of the sub-mm population. 
We also present the 850-$\mu$m SCUBA map of
this source. Throughout this paper we have assumed
$H_0=65$\,km\,s$^{-1}$\,Mpc$^{-1}$, $\Omega_M=0.3$ and $\Omega_\Lambda=0.7$. In
this cosmology the luminosity distance of J164502 is 11.2\,Gpc and $1\arcsecs$ 
corresponds to 9.1\,kpc.

\section{Observations and data reduction}
\subsection{Very Large Array observations}
At $z=1.44$ the \COJ{1}{0} line from J164502 falls within the VLA's Q-band at 0.7\,cm. 
The line widths of the \COJ{5}{4} and
\J{2}{1} lines were $\simeq 400$\,km\,s$^{-1}$ FWHM \citep{Andreani-et-al-2000}, and
a similar line width could be expected for the \J{1}{0} line.  
In order to cover the entire CO line, two adjacent 50\,MHz IF pairs (left and right
circular polarisation) were centred at 47.235 and 47.285\,GHz giving
635\,km\,s$^{-1}$ of coverage. 
Measurements of J164502 in the radio have yielded $S_{20\,cm} \leq 300\,\mu$Jy and
$S_{3.6\,cm} = (35 \pm 11)\,\mu$Jy \citep{Frayer-1996}. Fitting a power
law spectrum to these data points yields a spectral index of $\alpha \geq -1.25$.
Adopting a more conservative spectrum with $\alpha = -0.7$,
we find that the extrapolated synchroton flux at 0.7\,cm amounts to no more 
than $11\,\mu$Jy. Furthermore, the dust contributes by less than $10\,\mu$Jy at these
wavelengths \citep{Dey-et-al-1999}.
Hence, any contribution from continuum emission is neglible and no continuum
subtraction was necessary. Observations were obtained in D and C
configurations (see Table \ref{tbl-1}). Due to bad weather, all the C-configuration data
were of poor quality and discarded, leaving a total of 11.6\,hr of observing time on
source. The primary calibrator, 3C\,286, was observed at the beginning and end of
each observing run in order to fix the absolute flux density scale. Every 3-4 minutes the antennas were 
pointed toward a phase calibrator to ensure phase coherence throughout the run. This fast-switching 
technique resulted in a residual phase rms, after calibration, of less than $\sigma_{\phi} \sim 20\degs$.
Calibration and data reduction was done using standard recipes in the NRAO
$\mathcal{AIPS}$ Cookbook. D-configuration data taken at different times 
were combined using {\sc dbcon}, weighting each dataset by its 
total gridded weight.

The astrometrical uncertainty in the final CO map is given by $\sigma^2 = \sigma_{cal}^2
+\sigma_{res}^2 + \sigma_{SNR}^2$, where $\sigma_{cal}$ is the uncertainty of the assumed
position of the phase calibrator, $\sigma_{res}$ is the residual rms phase error 
after the phase correction, and $\sigma_{SNR}$ is the positional error due to the
presence of thermal noise.
In our case the latter is by far the most dominant source of error and
is given by $\sigma_{SNR} \simeq (FWHM/2)/SNR$. Adopting $FWHM\simeq 3\arcsecs$ 
and $SNR \simeq 7$ (see Fig.~1) we find the total positional error in the
final CO map to be $\sim 0.2\arcsecs$.

\subsection{Submillimetre observations}

Imaging data at 450 and 850-$\mu$m were obtained for J164502
using the Submillimetre Common User Bolometer Array (SCUBA --
\citet{Holland-et-al-1999}) during periods of exceptionally low opacity at the
James Clerk Maxwell Telescope (JCMT\footnote{The JCMT is operated by the Joint
Astronomy Centre on behalf of the United Kingdom Particle Physics and Astronomy
Research Council (PPARC), the Netherlands Organisation for Scientific Research,
and the National Research Council of Canada.}) on Mauna Kea, Hawaii.

Data for J164502 were acquired during 1998 July 28--30 and 1998 September
03--05, a total integration time of 28.8\,ks using a 45$''$ east-west
nod/chop. In brief, data were flat-fielded, corrected for
atmospheric losses, edited both automatically and by hand, plotted onto an
astrometric grid and flux calibrated using observations of planets and
secondary calibration sources. The map was convolved with a 
Gaussian resulting in a resolution of $14\arcsecs$.

\subsection{Near-IR Observations}
Near-IR K-band imaging was obtained of J164502 in June 2002 using the United Kingdom IR Telescope's
(UKIRT\footnote{UKIRT is operated by the Joint Astronomy Centre on behalf of the United
Kingdom Particle Physics and Astronomy Research Council (PPARC)}) UFTI imager, a
1024$^2$ HgTeCd array with $0.091\arcsecs$ pixels. Nine individual frames taken during
good to moderate seeing conditions were co-registered and
combined yielding a total exposure time of 2700s. The seeing in the final image is $\sim 0.8\arcsecs$.

Using 5 stars also detected in the USNO A2.0 catalogue \citep{Monet-et-al-1998},
the UKIRT image was calibrated on to the USNO reference frame to within $0.2\arcsecs$ rms.
The CO map was aligned to the K-band frame using the tasks {\sc geo} and {\sc hgeom}
in $\mathcal{AIPS}$ resulting in a relative positional accuracy between the two 
of $\sim 0.3\arcsecs$.

\section{Results}
In order to sensitively search for possible faint extended \COJ{1}{0} emission, 
at the expense of high spatial
resolution, we used the compact D-configuration data tapered
down using a Gaussian with a full-width of 60\,k$\lambda$ at 30 per-cent amplitude,
yielding a synthesized beam of $3.1\arcsecs \times 2.8\arcsecs$. 
The resulting CO map overlaid on the UKIRT K-band image is shown in Fig.~1.
\COJ{1}{0} emission is detected at the $7\sigma$ significance level in J164502. 
Assuming that we cover the entire \J{1}{0} line with the two IFs, and that we have
rectangular passbands, we estimate the velocity-integrated flux-density using
\begin{equation}
\int_{\Delta v} S_{\nu_{obs}} dv = 2 \Delta \nu_{IF} \frac{c}{\nu_{obs}} \overline{S}_{CO},
\end{equation}
where $\Delta \nu_{IF}$ is the width of the IF and $\overline{S}_{CO} = (S_1 + S_2)/2$ is the average
of the flux density measured in the two IFs. Using the task {\sc imean} to get the
flux density of the source in the two IF maps we find
$\overline{S}_{CO} = 1.0\pm0.2$\,mJy which yields a velocity-integrated flux density of
$0.6\pm 0.1$\,Jy\,km\,s$^{-1}$, where we have
used $\nu_{obs} = 47.26$\,GHz and an IF bandwidth of 45\,MHz instead of 50\,MHz due to bandpass rollover.

As first noted by \citet{Dey-et-al-1999} using high-resolution HST WFPC2 images, 
J164502 in the optical has a reflected S-shaped morphology with two bright knots at each end.
A comparison with Keck K-band imaging revealed that the bulk of the
near-IR emission comes from the region between the two bright knots located 
$\sim 0.4\arcsecs$ north of the brightest optical knot \citep{Dey-et-al-1999}.
As seen from Fig.~1, the peak intensity in the \COJ{1}{0} map coincides with the
peak in the UKIRT K-band image both of which are coincident with the
region of low optical emission found by \citet{Dey-et-al-1999}.
A similar offset between the HST position and the
\COJ{2}{1} emission was reported by \citet{Andreani-et-al-2000}.
 While the near-IR emission traces the dust-enshrouded stars 
in J164502, the optical emission corresponds to regions of low extinction, and it is therefore
not surprising to see such an offset between the two types of emission.
Since gas should be a good tracer of dust one would expect the CO to coincide with the
K-band emission, as it appears to do within the limits of our astrometric errors.

A Gaussian fit to the CO data in the image plane yields
a deconvolved source size of $4.5\arcsecs \times 3.0\arcsecs$, suggesting that
even in the D-configuration we have resolved the CO emission, albeit marginally.
Different weighting schemes and tapering functions were tried
in order to see the effects on the extended emission.
Using a less restrictive tapering function (FWHM$=200\rm{k}\lambda$) and
a robust parameter of 2 instead of 5 (natural weighting), 
thereby increasing the weight of the long baselines, the source still appeared 
extended although at a lower significance due
to the larger noise on the longest baselines (see Figure 1).  
In order to confirm or dismiss the reality of the extended emission
we shifted the peak of the CO emission
to the phase center, and then plotted the binned real and imaginary parts of
the complex visibilities as a function of baseline distance, 
Fig.~1. The complex visibilities are the Fourier components of
the source brightness distribution. Hence, for a Gaussian source 
brightness distribution the real part of the complex visibilities is
a Gaussian while the imaginary part will be zero since a Gaussian is
an even function. The dashed curves in Fig.~1 represents a least-squares 
fit of a circular Gaussian to the visibilities. The position of the Gaussian
was fixed to the phase center during the fit and only the amplitude and width 
were allowed to vary. We find that the real part
of the visibilities is consistent with a Gaussian source model with 
a $FWHM = 3.9\arcsecs$ in agreement with the Gaussian fit in the image plane.
The imaginary part is consistent with zero at all baselines within the limits
of our residual rms phase error of $\sigma_{\phi} \sim 20\degs$. The phase
error would give rise to a dispersion around zero of $\sigma(Im(V)) = 
\sigma(S \sin \phi) \simeq  S \sigma (\phi) = $1.0\,mJy $(20\degs/180\degs)\pi \simeq 
0.3$\,mJy which is close to the observed scatter.

The 850-$\mu$m SCUBA map of J164502 is shown in Fig.~2, and is seen to 
contain at least two statistically significant
sources. A $7\sigma$ emission feature showing in the map 
is detected at a position which coincides with the near-IR and CO-detections of J164502, 
and we take this to unambiguously be the 850-$\mu$m SCUBA detection of J164502.
Another source, marked as J164502-SMM1 in Fig.~2, detected at the $5\sigma$ level 
is seen $50\arcsecs$ to the south, with faint emission at the $4\sigma$ level
also seen just $\sim 20\arcsecs$ south of J164502. 
We estimated the 850-$\mu$m flux density in four different ways.
Two estimates were made by measuring the flux within apertures of $20\arcsecs$ 
and $16\arcsecs$ diameter. A third measurement was made by fitting a Gaussian to the
emission, and a fourth by fitting a Gaussian with a fixed $FWHM$ of $15\arcsecs$
to the emission. The values derived from each of these measurements are given
in Table \ref{tbl-2} for J164502 as well as J164502-SMM1. For J164502 we find
$S_{850\mu\rm{m}} = 8 \pm 2$\,mJy where we have adopted 
the average of the four different measurements. While this is in good agreement
with the 850-$\mu$m measurement by \citet{Cimatti-et-al-1998}, it is larger than the 
flux quoted by \citet{Dey-et-al-1999}. Both of these measurements were obtained using SCUBA in 
its photometry mode but the latter value is based on twice as large a data-set as the former,
taken in excellent weather conditions and is therefore the most reliable estimate. 
If this is the case, the large discrepancy between the flux density derived from our map
and the value derived from photometry-measurement by \citet{Dey-et-al-1999} could be due 
to extended emission from J164502 which would have been missed by the single pixel 
measurement.

From the 850-$\mu$m map in Fig.~2 it is seen that the emission from J164502 does in fact appear to be extended,
with emission to the NE. The reality of this extended emission is strengthened by the high-resolution Keck K-band
image obtained by \citet{Dey-et-al-1999} which shows an emission feature extending in the same direction, albeit on
a smaller scale of $\ga 1\arcsecs$ (see Fig.~2 of \citet{Dey-et-al-1999}). However, the fact that the stars traced by the K-band
emission is extended on scales of $\ga$9\,kpc means that cold dust is distributed
on at least the same scales, and likely on much larger scales. 
Fitting a Gaussian to the SCUBA image yields a source size of $15.1\arcsecs$ along the major axis,
which is comparable to the $14\arcsecs$ resolution of the map, though very 
tentatively suggestive of a slight extension.
In Fig.~2 we compare the azimuthally averaged radial
profiles of J164502 and J164502-SMM1 with the point spread function (PSF) of SCUBA at 850-$\mu$m 
derived from a beam map of a bright blazar obtained during the same observing run. 
The radial profile of J164502 agrees with the PSF out to $r\simeq 20\arcsecs$ beyond
which it displays excess emission over that of a point source. This bump
in the radial profile is due to the $4\sigma$ emission feature $20\arcsecs$ south ($\sim 180$\,kpc) of
J164502. The reality of this emission feature is questionable but if real 
its close vicinity to J164502 could mean that it is a system in the process of
merging with J164502.
While there is thus tentative evidence that J164502 is extended, the large beam size of
SCUBA does not allow for a firm conclusion on this issue.

For comparison we note that J164502-SMM1 is consistent with the PSF at all radii.
The SCUBA 850-$\mu$m beam is too large to make a reliable identification of J164502-SMM1 with an 
near-IR/optical source. If J164502-SMM1 is a dusty system at the same redshift
as J164502 one might expect that it too would shine in CO. No CO emission
was detected at the position of J164502-SMM1, however, and
we set an upper limit on the velocity-integrated line flux using the equation
$S_{\rm{CO}}\Delta v < 3 \sigma (\delta v dv)^{1/2}$,
where $\sigma$ is the rms noise,  $\delta v$ the velocity resolution, and
$dv$ the line width which we assume to be equal to the \J{2}{1} 
line width, i.e. $\sim 400$\,km\,s$^{-1}$ \citep{Andreani-et-al-2000}. The
measured rms noise in the image is $\sigma \simeq 0.1$\,mJy\,beam$^{-1}$
which yields $S_{\rm{CO}}\Delta v < 0.2$\,Jy\,km\,s$^{-1}$ for a velocity-
coverage of $635$\,km\,s$^{-1}$. The failure to detect CO-emission from J164502-SMM1
does not rule out a association since cluster velocity dispersions can go up to 
$\ga 1000$\,km\,s$^{-1}$, and in some cases even closely associated CO-emitting regions can have velocity
differences of $\sim 1000$\,km\,s$^{-1}$ as was seen in the high redshift
radio galaxy 4C\,60.07 \citep{Papadopoulos-et-al-2000}. The probability of finding by chance
a source with a flux density brighter than or equal to 6\,mJy within $r=50\arcsecs$ from 
J164502 is given by  
$P = 1 - e^{-\pi r^2 N}$, where $N$ is the average surface density of sources with 
$S_{850\mu\rm{m}} \ge 6$\,mJy \citep{Downes-et-al-1986}. Adopting a surface density of 
$N\sim 400$\,deg$^{-2}$ \citep{Borys-et-al-2003} one
finds that $P = 0.22$. On that basis we conclude that
J164502-SMM1 is most likely unrelated to J164502.

\section{Analysis \& Discussion}
Table \ref{tbl-2} summarises the CO and sub-mm observations of J164502 
and J164502-SMM1 as well as the physical quantities derived from them.

\subsection{Molecular gas and dust in J164502}
For an observed velocity-integrated line flux density of the \JJ{J+1}{J} CO
line, the intrinsic CO line luminosity is given by
\begin{equation}
\rm{L}_{\rm{CO}_{J+1,J}}=\frac{c^2}{2k \nu^2_{J+1,J}} 
\frac{D^2_L}{1+z} \int_{\Delta v} S_{\nu_{\rm{obs}}} dv,
\end{equation}
where $D_L$ is the luminosity distance, and $\nu_{J+1,J}$ is the rest-frame
frequency of the CO transition. Inserting astrophysical units, the equation
reads
\begin{equation} 
\rm{L'}_{\rm{CO}_{J+1,J}}=2.43\times 10^{3} (J+1)^{-2} 
(1+z)^{-1}\left ( \frac{D_L}{\rm{Mpc}}
\right )^2 \left ( \frac{\int_{\Delta v} S_{\nu_{\rm{obs}}} dv}{\rm{Jy}\, \rm{km}\, 
\rm{s}^{-1}} \right ),
\label{L-CO}
\end{equation} 
where $\rm{L'}_{\rm{CO}_{J+1,J}}$ is the so-called pseudo-luminosity which is
measured in units of K\,km\,s$^{-1}$\,pc$^2$.  
The observed \COJ{1}{0} line flux density for J164502 implies
an intrinsic CO luminosity of $\LOJ{1}{0} = (7 \pm 1)\times 
10^{10}$\,K\,km\,s$^{-1}$\,pc$^2$.
From the \COJ{1}{0} line the molecular gas mass can be found using the scaling
relation $M(\rm{H}_2) = X_{CO} \LOJ{1}{0}$ where 
$\XCO\simeq 5\, \Msolar\,(\rm{K}\,\rm{km}\,\rm{s}^{-1}\,\rm{pc}^2)^{-1}$ is
the standard Galactic CO-H$_2$ conversion factor. The standard conversion
factor, \XCO~has been calibrated using Giant Molecular Clouds (GMCs) in our
Galaxy \citep{Strong-et-al-1988}. However, the conditions of the interstellar medium in active
starbursting galaxies such as ULIRGs is markedly different from that in our
Galaxy, and applying the Galactic conversion factor would result in seriously
underestimated gas masses. In particular, the concept of isolated and
virialised GMCs breaks down, and instead the bulk of the CO emission is found
to originate from a warm diffuse phase. In effect, this yields a lower conversion 
factor of $\XCO=0.8\, \Msolar\,(\rm{K}\,\rm{km}\,\rm{s}^{-1}\,\rm{pc}^2)^{-1}$, more
appropriate for such extreme environments \citep{Downes-and-Solomon-1998}.  
It is reasonable to assume that in high-redshift starbursts similar conditions
prevail and the aforementioned value of \XCO~is also adopted for J164502.
In doing so we estimate the molecular gas mass present in J164502
to be $M(\rm{H}_2) =  (6\pm 1) \times 10^{10}\,\Msolar$. 
This is in agreement with the \J{2}{1} detection by \cite{Andreani-et-al-2000}, which 
yields a molecular gas mass of $\sim 4 \times 10^{10}\,\Msolar$ when converted
to the cosmology adopted in this paper.

The \COJ{1}{0} line luminosity and the implied molecular gas mass of J164502 is 
somewhat lower than those found in the quasars APM\,08279+5255 and PSS\,2322+1944, and 
in the high redshift radio galaxy 4C\,60.07
($\LOJ{1}{0} \sim 10^{11}$\,K\,km\,s$^{-1}$\,pc$^2$ and $M(\rm{H}_2) \sim 10^{11}\,\Msolar$ -
see Papadopoulos et al.~2001; Carilli et al.~2002; Greve et al.~in preparation). This is not
surprising since these objects
are among the most luminous systems in the Universe,
and might therefore be expected to contain more gas.
\citet{Solomon-et-al-1997} observed \COJ{1}{0} in a sample of 37 local ULIRGs out to
a redshift of $z=0.3$, and found an average \J{1}{0} luminosity
of $\LOJ{1}{0} \simeq 8 \times 10^9$\,K\,km\,s$^{-1}$\,pc$^2$ which is roughly
the luminosity of Arp\,220 and Mrk\,231. The scatter on this result was only
30 per cent. A more recent survey
of \COJ{1}{0} observations of a complete sample of 60 
ULIRGs selected from the SCUBA Local Universe Galaxy Survey (SLUGS - \citet{Dunne-et-al-2000})
yielded similar results \citep{Yao-et-al-2003}. Hence, we find that J164502 has a \COJ{1}{0}
luminosity and consequently a molecular gas mass which is about an order of magnitude 
larger than is found for local ULIRGs.

The dynamical mass of the system can be calculated from the observed size and
line width of the source and assuming the gas is distributed in a disk with
diameter $L$ in Keplerian rotation, in which case it can be shown that the dynamical mass
is given by 
\begin{eqnarray}
M_{dyn} &\simeq& \frac{\Delta V_{FWHM}^2 L}{2 \alpha_d G} \frac{1}{\sin^2 i}  \\
      &   =  & 1.16\times 10^9 \left ( \frac{\Delta V_{FWHM}}{100\rm{\,km\,s}^{-1}} 
              \right )^2 \times \left (\frac{R}{\rm{kpc}}\right )\frac{1}{\sin ^2 i} 
                 \Msolar,
\end{eqnarray}
where $i$ is the inclination angle of the disk, and $\alpha_d$ is a correction
factor of order unity \citep{Bryant-Scoville-1996}. 
Adopting a line width of $\Delta V_{FWHM} = 400$\,km\,s$^{-1}$ \citep{Andreani-et-al-2000}, a 
source size of $\theta \la \sqrt{4.5\arcsecs \times 3.0\arcsecs} = 3.7\arcsecs$, 
which corresponds to a maximum linear diameter of $L \sim 34$\,kpc at a redshift of $z=1.44$
we estimate the enclosed dynamical mass within the CO-emitting region to be
$M_{dyn} \la 6.3 \times 10^{11}(\sin i)^{-2}\,\Msolar$. 
The inferred ratio of the molecular-to-dynamic mass for
the system as a whole is then $M(\rm{H}_2)/M_{dyn} \ga 0.1 \sin ^2 i$.
Hence, geometrical factors aside, the amount of molecular gas accounts for at least 
10 per cent of the total dynamical mass within the CO-emitting region. 
The dynamical mass should be considered an upper limit on the total amount of molecular gas
present in J164502, and the two mass estimates would only coincide if the CO emission was concentrated
within the inner $\sim 3$\,kpc.
Due to the higher spatial resolution and the fact that the \J{1}{0} line provides
an unbiased estimate of $M(\rm{H}_2)$, our constraint on the molecular-to-dynamical
mass ratio should be an improvement over that of \citet{Andreani-et-al-2000}.
A gas mass fraction of $\ga 10$ per cent is compatible with the range of gas mass fractions 
found for local ULIRGs \citep{Downes-and-Solomon-1998} yet, unlike the latter, it is 
distributed over significantly larger scales.

The extrapolated synchroton radiation flux at 850-$\mu$m is less than $3\,\mu$Jy, and it is therefore 
safe to assume that 
the observed 850-$\mu$m flux from J164502 is dominated
by thermal dust emission, with the radio synchroton emission contributing
a negligible amount. The dust mass can therefore be estimated
using
\begin{equation}
M_d = \frac{S_{\nu_{\rm{obs}}} D_L^2}{(1+z) \kappa_d(\nu_{\rm{rest}})} 
\left [B(\nu_{\rm{rest}},T_d) - B(\nu_{\rm{rest}},T_{cmb}(z) \right ]^{-1},
\end{equation}
where $\nu_{rest} = \nu_{obs} (1 + z)$ is the rest-frame frequency, $T_{cmb}(z)$ is the CMB temperature
at redshift $z$, and $\kappa_d \propto \nu^\beta$ is the dust absorption coefficient.
The emissivity index, $\beta$, depends on the dust temperature, $T_d$, 
\citep{Dunne-et-al-2000} both of which are 
poorly constrained for high-redshift objects - typical values are $\beta = 1-2$.  
\citet{Dey-et-al-1999} found the spectral energy distribution (SED) of J164502 to be
well described by an optically thin modified black body law with a dust 
temperature of $T_d=40$\,K and an emissivity law of $\beta=1.5$, and we
adopt those values here.
In doing so we estimate the dust mass to be 
$M_{dust}\simeq 9 \times 10^8\,\Msolar$, where we
have used a dust absorption coefficient of 
$\kappa_d(\nu_{rest}) = 0.11 (\nu_{rest}/353\,\rm{GHz})^\beta$\,m$^2$\,kg$^{-1}$ after
\cite{Hildebrand-1983}. The uncertainty in $k_d$ is large, so we present this estimate
for comparison only.
We can estimate the total far-IR luminosity by integrating
the thermal spectrum
\begin{equation}
L_{FIR} = 4 \pi M_d \int_{0}^{\infty} \kappa_d (\nu) B(\nu, T_d) d\nu,
\end{equation}
which yields a total far-IR luminosity of $L_{FIR} \simeq 9 \times 10^{12}\,\Lsolar$. 
This almost puts J164502 in the class of hyperluminous infrared galaxies 
(HyLIRGs) which have $L_{FIR} \ga 10^{13}\,\Lsolar$.  
The dust masses and far-IR luminosities typically found in local ULIRGs are 
$M_{dust} \simeq 10^{7-8}\,\Msolar$ and $L_{FIR} \ga 10^{12}\,\Lsolar$
\citep{Sanders-and-Mirabel-1996,Dunne-et-al-2000}, i.e.~nearly an order of magnitude
smaller than what we find for J164502. It is comparable, however, to what is found in 
high-redshift QSOs and HzRGs \citep{Omont-et-al-2003,Archibald-et-al-2001} and,
perhaps more importantly, similar to the dust masses and far-IR luminosities
found in sub-mm-selected dust-enshrouded starbursts at high redshifts 
(Ivison et al.~1998, 2000).
From the above we estimate the gas-to-dust mass ratio of
J164502 to be $M(\rm{H}_2)/M_{dust} \simeq 67$
which is well within the range of 50-100 found for local ULIRGs.
Note, however, that significant uncertainty is attached to the
normalisation of $\kappa_d$, and values differing by as much as a factor of two has
been reported \citep{Draine-and-Lee-1984,Mathis-and-Whiffen-1989}, which in 
turn could lead to a similar change in the dust mass but not in $L_{FIR}$ 
which is independent of the normalisation value of the dust absorption coefficient. 
Using $\beta=2.0$ instead of 1.5 would decrease the dust mass by nearly a 
factor of two, but increase $L_{FIR}$ by a similar amount.

If the bulk of the far-IR emission can be ascribed to starburst
activity the corresponding star formation rate is given by
\begin{equation}
SFR \simeq L_{FIR} 10^{-10}\delta_{IMF} \delta_{SB} \,\Lsolar\,\Msolar\,\rm{yr}^{-1},
\end{equation}
where $\delta_{IMF} \sim 1-6$ is a function of the initial mass function, and
$\delta_{SB}$ is the fraction of the FIR emission which is heated by the starburst
\citep{Omont-et-al-2001}. Assuming a conservative value of $\delta_{IMF} = 1$, we
estimate the starformation rate in J164502 to be $SFR\simeq 900\delta_{SB}\,\Msolar\,\rm{yr} ^{-1}$. 
It is possible that J164502 harbours an AGN in its center, so
a significant fraction of the far-IR
luminosity could be due to dust being heated by the AGN and not by the
starburst. The derived value for the SFR could therefore be 
overestimated. However, the narrow linewidths seen in optical and NIR spectra
of J164502 \citep{Dey-et-al-1999} favours young hot
stars over an AGN as the main source of energy. Furthermore, as pointed
out by \citet{Dey-et-al-1999}, J164502
deviates from the 60\,$\mu$m - 6\,cm relation obeyed by local star-forming galaxies
by having almost an order of magnitude less radio emission at 6\,cm, which is
unlikely if an AGN dominated the energetics of the system \citep{Dey-and-van-Breugel-1994}.
In addition, it is found 
that the bulk of the far-IR emission from local ULIRGs is powered by starburst 
activity even though an AGN is
present \citep{Downes-and-Solomon-1998}. Finally, if the 
sub-mm emission is extended it would strongly suggest that the main power source for the
far-IR emission is a massive starburst since it is difficult to imagine
the AGN heating the dust on scales of tens of kpc. The large quantity of
molecular gas revealed by the CO detections in this system could provide
the necessary fuel for such a massive starburst for a period of $\sim 10^7$ years.

The efficiency of star formation should be measured relative to the amount of
molecular gas available to form stars.  Such a measure of the star-formation
efficiency is the rate of star formation per solar mass of molecular hydrogen,
i.e. $L_{FIR}/M(\rm{H}_2)$. For J164502 we find $L_{FIR}/M(\rm{H}_2)\simeq
150\,\Lsolar\,\Msolar^{-1}$. Probably a better gauge of the star-formation efficiency is
the continuum-to-line ratio $L_{FIR}/L'_{\rm{CO}}(1-0)$ since it is 
independent of \XCO. Giant Molecular Cloud
(GMCs) in our Galaxy typically have values of $\sim 15$ and similar ratios are
found in nearby spirals \citep{Mooney-and-Solomon-1988}. Starburst galaxies and
ULIRGs have $L_{FIR}/L'_{\rm{CO}}(1-0)$ ratios which are 10 times higher than this, ranging
in values from 80 to 250 with a median of
$160\,\Lsolar($\,K\,km\,s$^{-1}$\,pc$^2)^{-1}$ \citep{Solomon-et-al-1997}. 
For J164502 we find $L_{FIR}/L'_{\rm{CO}}(1-0)\simeq 129\,\Lsolar
($\,K\,km\,s$^{-1}$\,pc$^2)^{-1}$, i.e.~in line with what is found for
local ULIRGs.
Recent studies of large samples of ULIRGs at low and intermediate redshifts 
have shown that the $L_{FIR}/L'_{\rm{CO}}(1-0)$
ratio increases with increasing $L_{FIR}$ \citep{Young-et-al-1986,Tutui-et-al-2000,Yao-et-al-2003}.
Such behaviour can be explained if the more far-IR luminous galaxies, in addition 
to having more dust and gas which would just continue the linear relation between
$L_{FIR}$ and $L'_{\rm{CO}}(1-0)$, have higher dust temperatures 
due to extra heating by an AGN or a higher star-formation efficiency.
Assuming that there is no contribution to $L_{FIR}$ from an AGN, the derived
$L_{FIR}/L'_{\rm{CO}}(1-0)$-ratio should probably be taken
as  a lower limit on the star-formation efficiency 
since in the extreme starburst regions dominating the emission from ULIRGs, the bulk of the CO luminosity
comes from a diffuse inter-cloud medium, rather than from the dense gas
gravitationally bound in clouds where the stars are formed
\citep{Downes-et-al-1993,Solomon-et-al-1997}. The dense gas is better traced by
HCN, and as shown by \citet{Gao-et-al-1999} the $L_{FIR}/L_{\rm{HCN}}$-ratio is the
same for GMCs to ULIRGs, indicating that anywhere in the Universe only the dense
gas is relevant to star-formation. In ULIRGs the dense/diffuse gas mass ratio
(roughly quantified by $L_{\rm{HCN}}/L_{\rm{CO}}$) is particularly high which is not
a surprise given their merger status and the way the gas responds to a merger.

\subsection{Excitation conditions of the molecular gas in J164502}
The velocity/area-averaged brightness temperature ratio between CO \JJ{J+1}{J} and CO \JJ{1}{0}
is
\begin{equation}
r_{J+1,J} = \frac{T_b(J+1\rightarrow J)}{T_b(1\rightarrow 0)} = \frac{L'(J+1\rightarrow J)}{L'(1\rightarrow 0)}.
\end{equation}
\citet{Andreani-et-al-2000} estimated the \COJ{2}{1} and \J{5}{4} luminosities
to be $4\times 10^{10}$ and $7\times 10^9$\,K\,km\,s$^{-1}$\,pc$^2$, respectively,
where we have converted the luminosities to the cosmology adopted here.
The line ratios estimated from eq.~8 are
$r_{21} = 0.6\pm 0.2$ and $r_{54} = 0.10\pm 0.05$.
The majority of high-redshift systems detected in CO to date are strongly gravitationally
lensed. This makes estimating the excitation conditions in such systems
complicated since differential magnification of the high-J lines compared to lower
transitions may bias the line ratios significantly.
There is nothing to suggest that J164502 is gravitationally lensed, and the above
derived line ratios ought to represent the intrinsic excitation conditions
of the gas, albeit still averaged over the entire galaxy. 

We used the observed line ratios of $r_{21}=0.6\pm  0.2$ and
$r_{54}=0.10\pm 0.05$ as constraints to a Large Velocity Gradient
(LVG) code in order to gain some insight on the bulk properties of the
molecular gas in J164502. A lower limit on gas temperature 
$\rm T_{kin}=40$\,K is assumed since this value is deduced for the bulk of
the dust \citep{Dey-et-al-1999} and $\rm T_{kin}\ga T_{dust}$  
is expected for FUV-heated gas/dust \citep{Tielens-and-Hollenbach-1999}. 
A wide range of temperatures  
($\rm T_{kin}=60-90$\,K) offers an acceptable fit to the aforementioned
line ratios, with $\rm  T_{kin}=70$\,K being the optimal value. This
wide range is due to the poor constraints offered by only two
line ratios while the high gas temperatures 
may reflect the UV-intense environment of the molecular gas in J164502
and/or turbulent motions heating only the gas, a situation that has
already been noted for molecular gas in the Galactic center
\citep{Rodriguez-Fernandez-et-al-2001}.

A common  feature of all  the LVG solutions  is the low  gas densities
$\rm n(H_2)\sim 300\  cm^{-3}$. This has already been  noted when only
the CO $(5-4)/(2-1)$ line  ratio was available \citep{Papadopoulos-and-Ivison-2002}
and is a property often found for the bulk of the gas in extreme
starbursts  \citep{Aalto-et-al-1995,Downes-and-Solomon-1998}.
Interestingly  for such low  gas densities  all CO \JJ{J+1}{J}
transitions with $J+1>5$ have flux  density ratios with respect to the
lowest $1-0$ transition  of $S(J+1-J)/S(1-0)=(J+1) ^2\ r_{\rm J+1\
J}\la  0.25$  for  any  plausible gas  temperature  ($\rm  T_{kin}\sim
40-100$\,K), thus  no longer offering  the advantage of a higher
flux density with respect to the latter.

As a consequence, any estimate of the H$_2$ mass based on the assumption
of optically thick, thermalised high-J CO lines could be severely underestimated.
Using the 5--4 line in J164502 only, and assuming it is optically thick and
thermalised ($r_{54} \sim 1$), one would find a molecular gas mass which is an order
of magnitude lower than that inferred from the \J{1}{0} line.
There is no significant flux advantage between the 2--1 and 1--0 lines, since
$S_{\rm{CO}}(2-1)/S_{\rm{CO}}(1-0) = 4 T_b(2-1)/T_b(1-0) > 1$ even
for a very subthermal ratio of $r_{21} = 0.3$, say. This is consistent with the
fact that we find the molecular masses derived from the two lines 
to be similar (section 4.1). 
The above demonstrates the importance of using low-J CO lines in order to
infer the amount of molecular gas in a galaxy. This is particularly true 
at redshifts beyond 3, since at those redshifts the current mm-interferometers can
only hope to detect CO \JJ{J+1}{J}, $J > 2$, and the the excitation bias of the high-J
lines could become very severe.

\subsection{Comparison with high-redshift sub-mm galaxies}
Although J164502 was originally selected as an ERO and not a (sub)millimetre
galaxy, subsequent observations at optical/IR and sub-mm wavelengths have shown
that it can in fact be considered a typical sub-mm galaxy
(Dey et al.~1999; see also Smail et al.~1999). Our detection of
\COJ{1}{0} in J164502 is therefore the first detection of this transition in a galaxy
which is thought to be similar to the SCUBA population of dust-enshrouded
galaxies at high redshifts. 
J164502 has a far-IR and CO luminosity about ten times that of the
average values found in ULIRGs, and the amount of molecular gas present in J164502 is comparable to
the median gas mass ($\sim 2 \times 10^{10}\,\Msolar$) of the five SCUBA galaxies 
detected in CO to date \citep{Neri-et-al-2003}, although a thorough
characterisation of the gas content of this population has to await 
CO observations of a large, unbiased sample of submm-selected galaxies. 
Hence, the observations presented here further supports that J164502 is more similar in
its properties to sub-mm galaxies than a high-redshift analogue of Arp\,220.

Recently, a subject of some debate has been whether the CO emission detected
from high-redshift sub-mm galaxies originates from a huge, massive resevoir
of molecular gas $\sim 10$\,kpc in size
or from a much more compact circumnuclear disk, typically of radius $r\sim 100$\,pc,
as seen in local ULIRGs such
as Arp\,220 \citep{Ivison-et-al-2001,Genzel-et-al-2003,Downes-and-Solomon-2003}. If, as the observations seem
to suggest, the starburst in J164502 is extended over several tens of kpc
that would set apart from the local ULIRG population and point toward
the former scenario.

\acknowledgments
TRG acknowledges support from the Danish Research Council and from the EU RTN
Network POE. PPP acknowledges a Marie Curie Individual Fellowship HPMT-CT-2000-00875. 
The authors wish to thank Ignas Snellen 
and Philip Best for useful advices on $\mathcal{AIPS}$, and Omar Almaini for 
kindly providing us with the reduced UKIRT UFTI K-band image of J164502+4626.4.
We also thank Jason Stevens and Andy Taylor for useful discussions.

\clearpage


\begin{figure}
\plottwo{f1.eps}{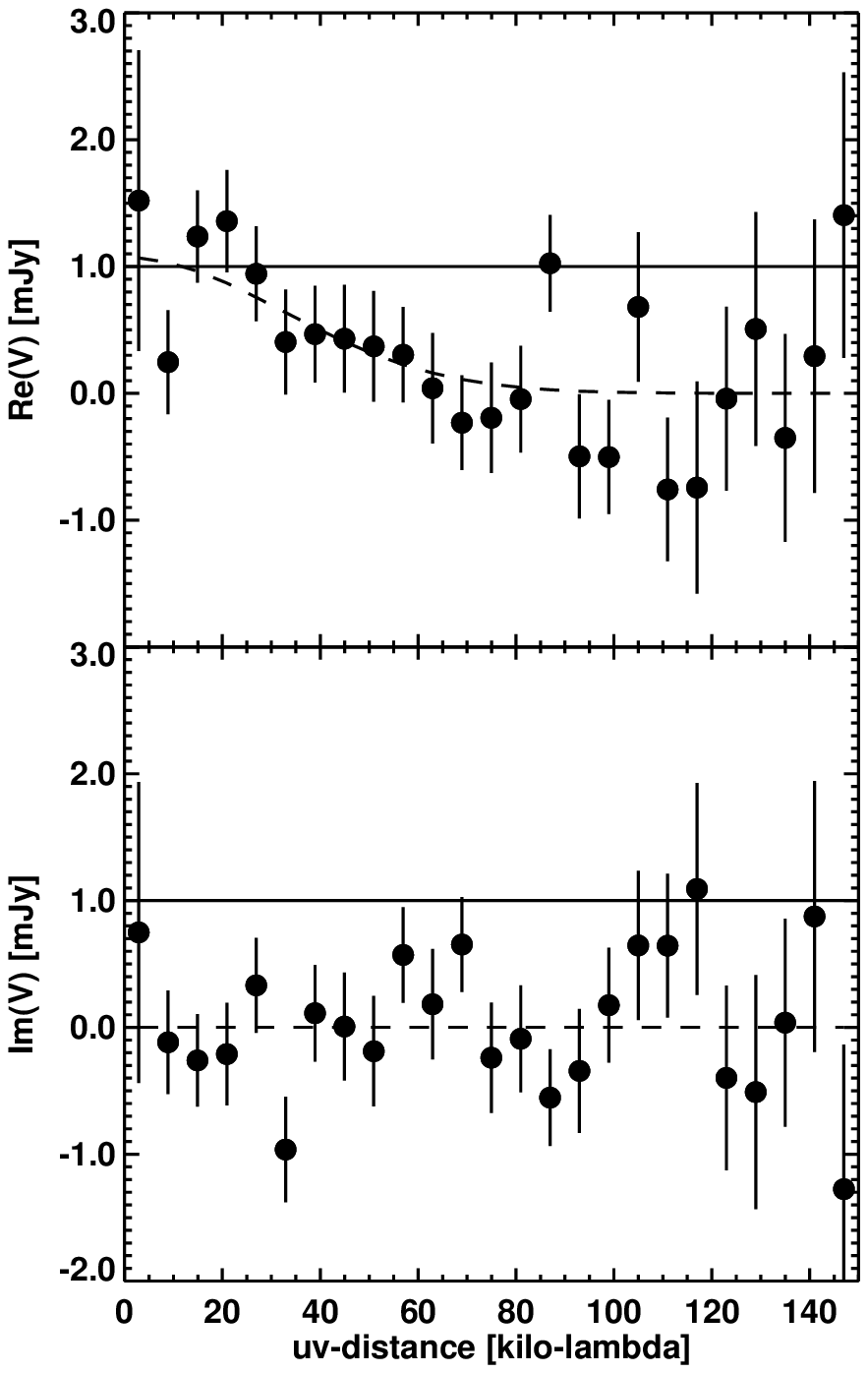}
\caption{\textbf{Left:} Naturally weighted \COJ{1}{0} contour map of J164502 overlaid on 
the UKIRT UFTI K-band image. The resolution of the CO map
is $3.1\arcsecs \times 2.8\arcsecs$ at $PA=157\degs$, see the insert.
The contours are -2,2,3,4,5,6, and $7\sigma$ where $\sigma = 0.1$\,mJy\,beam$^{-1}$.
\textbf{Right:} The real (top panel) and imaginary (bottom panel) part of 
the complex visibilities vs.~baseline. The visibilities have been binned. The solid line 
represents a point source with flux density of 1.0\,mJy. The dashed curves represent a 
least-square fit of a Gaussian source model to the visibilities.}
\end{figure}

\clearpage 

\begin{figure}
\plottwo{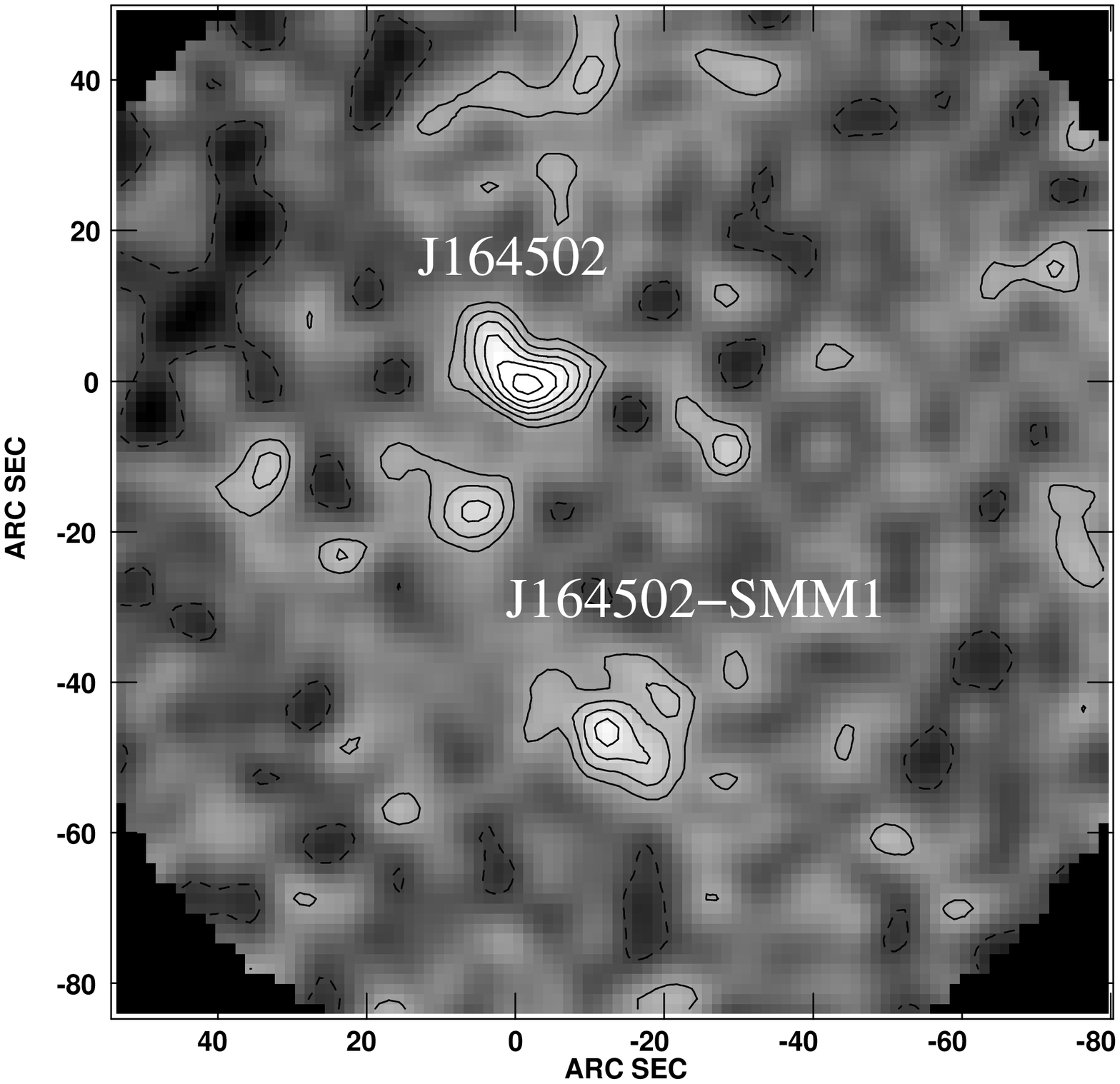}{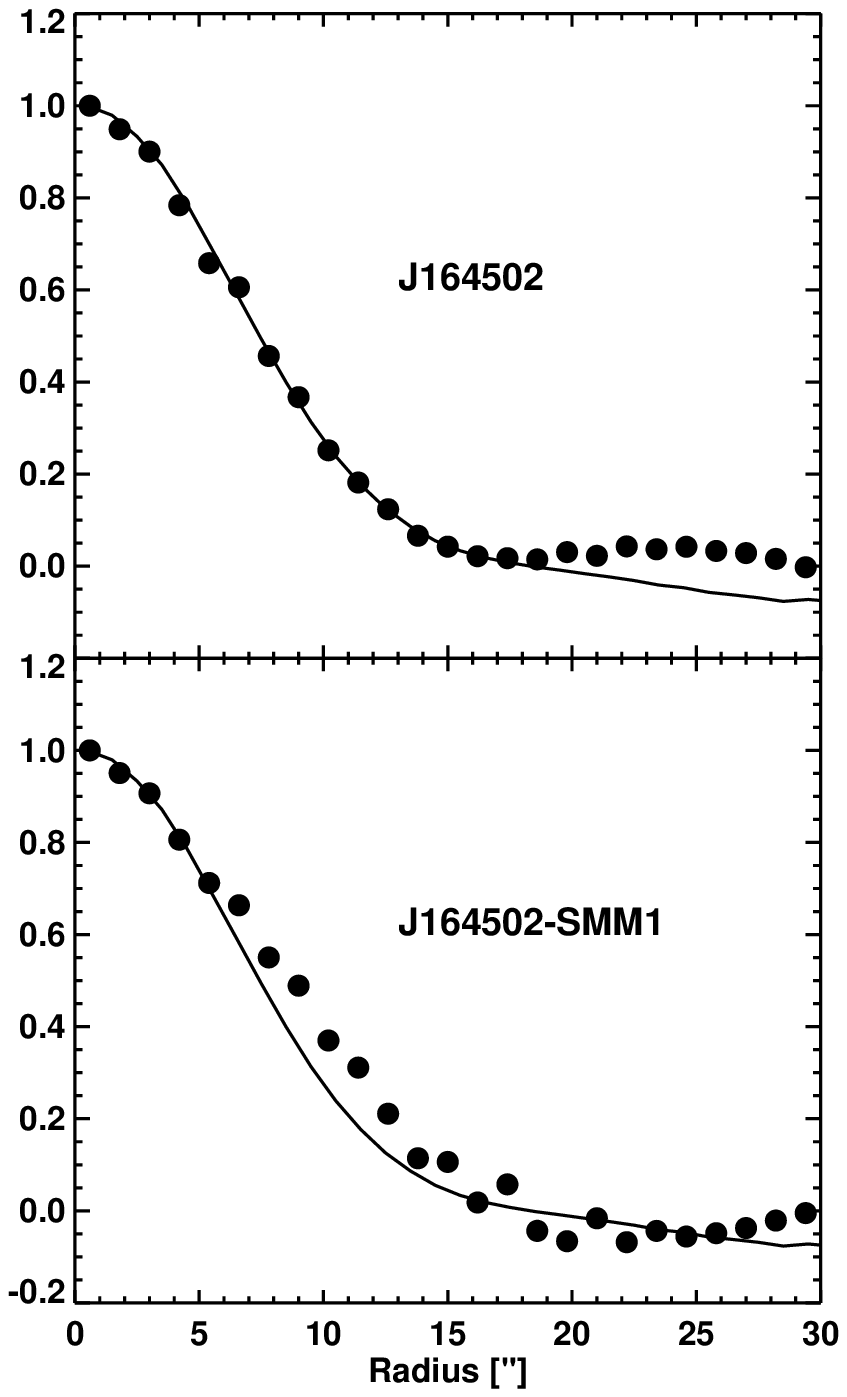}
\caption{\textbf{Left:} SCUBA 850-$\mu$m jiggle-maps of J164502. The
resolution is $FWHM = 14\arcsecs$, and the rms noise level is 1.6\,mJy\,beam$^{-1}$.
Contours start at $2\sigma$. J164502-SMM1 denotes a statistically significant detection
of 850-$\mu$m emission about $50\arcsecs$ south of J164502. Weak
emission is also seen $20\arcsecs$ south-east of J164502.
\textbf{Right:} Radial profiles of J164502 and J164502-SMM1 (filled circles) compared
with the radial profile of the PSF (solid curve).} 
\end{figure}

\clearpage






\clearpage
\begin{deluxetable}{lccc}
\tabletypesize{\scriptsize}
\tablecaption{VLA Observations \label{tbl-1}}
\tablewidth{0pt}
\tablehead{
\colhead{Date} & \colhead{Configuration}  & \colhead{Frequency, (GHz)} & Integration time, (hrs)}
\startdata
 2001 Oct  9     & D    & 47.235,47.285                    &  5.2\\
 2002 Oct 26     & C    & 47.235,47.285                    &  2.9\\
 2002 Nov 14     & C    & 47.235,47.285                    &  2.4\\
 2002 Dec 24     & C    & 47.235,47.285                    &  2.6\\
 2003 Feb 21     & D    & 47.235,47.285                    &  2.6\\
 2003 Mar  1     & D    & 47.235,47.285                    &  3.8\\
\enddata
\end{deluxetable}

\clearpage
\begin{deluxetable}{lccc}
\tabletypesize{\scriptsize}
\tablecaption{Observed properties of J164502 \label{tbl-2}}
\tablewidth{0pt}
\tablehead{
\colhead{Parameter}  & \colhead{J164502} & \colhead{J164502-SMM1} }
\startdata
$\alpha\, $(J2000)     & 16:45:02.26  & 16:45:01.20  \\
$\delta\, $(J2000)     & +46:26:26.50 & +46:25:39.0  \\
$z_{\rm{CO}}$          & 1.439        & \\
$\int_{\Delta V} S_{\rm{CO}(1-0)}dV$\,[Jy\,km\,s$^{-1}$] & $ 0.6\pm 0.1$ &  $<0.2$\\
$S_{850\mu\rm{m}}$\,[mJy]  & $8\pm 2$\tablenotemark{a}   &   $6 \pm 2$\tablenotemark{b}\\
$\LOJ{1}{0}$\,[K\,km\,s$^{-1}$\,pc$^{-1}$]  & $(7\pm 1)\times 10^{10}$  &  $<2 \times 10^{10}$\tablenotemark{c} \\
$L_{FIR}$\,[$\Lsolar$]  & $9\times 10^{12}$ & $7 \times 10^{12}$\tablenotemark{c}\\
\enddata


\tablenotetext{a}{The average of 7.5/7.9/7.3/9.8\,mJy which was the flux densities obtained using the four different methods described in section 3.}
\tablenotetext{b}{The average of 6.7/6.5/6.0/6.5\,mJy which was the flux densities obtained using the four different methods described in section 3.}
\tablenotetext{c}{This value assumes that J164502-SMM1 has a redshift of $z=1.439$.}

\end{deluxetable}


\end{document}